\documentclass[preprint]{aastex} 

\usepackage[dvips, bookmarks, colorlinks=true, pdftitle={Terrestrial
    Planet Formation from an Annulus}, pdf author={Kevin Walsh},
  pdfsubject={Terrestrial planet formation from an annulus},
  pdfkeywords={Planet formation, Annulus}]{hyperref}
\usepackage{graphicx} \usepackage{amsfonts} \newcommand{\hide}[1]{} %

\usepackage{wasysym}

\newcommand{\rinit}{$R_\mathrm{init}$}

\begin{document}
\title{Terrestrial Planet Formation from an Annulus}
\author{Kevin J. Walsh,  Harold F. Levison}
\affil{Southwest Research Institute, 1050 Walnut St. Suite 300, Boulder, CO 80302, USA}
\email{kwalsh@boulder.swri.edu}

\begin{abstract}

It has been shown that some aspects of the terrestrial planets can be
explained, particularly the Earth/Mars mass ratio, when they form from
a truncated disk with an outer edge near 1.0~au
\citep{Hansen:2009p8802}.  This has been previously modeled starting
from an intermediate stage of growth utilizing pre-formed planetary
embryos.  We present simulations that were designed to test this idea
by following the growth process from km-sized objects located between 0.7--1.0~au up to terrestrial planets.  The simulations explore 
initial conditions where the solids in the disk are planetesimals with
radii initially between 3 and 300~km, alternately including effects
from a dissipating gaseous solar nebula and collisional fragmentation.
We use a new Lagrangian code known as {\tt LIPAD}
\citep{Levison:2012p12338}, which is a particle-based code that models
the fragmentation, accretion and dynamical evolution of a large number
of planetesimals, and can model the entire growth process from
km-sizes up to planets.

A suite of large ($\sim$ Mars mass) planetary embryos is complete in
only $\sim$~1~Myr, containing most of the system mass. A quiescent
period then persists for 10-20~Myr characterized by slow diffusion of
the orbits and continued accretion of the remaining
planetesimals. This is interupted by an instability that leads to
embryos crossing orbits and embyro-embryo impacts that eventually
produce the final set of planets. While this evolution is different
than that found in other works exploring an annulus, the final
planetary systems are similar, with roughly the correct number of
planets and good Mars-analogs.

\end{abstract}


\section{Introduction}

An important and challenging issue in understanding terrestrial planet
formation are the large differences between Earth and Mars. The two
planets are solar system neighbors but are separated by an order of
magnitude in mass and accretion age
\citep{Nimmo:2007p11241,Kleine:2009p9784,Dauphas:2011p19768}. Classical
models of terrestrial planet formation with initial conditions that
include smooth disks of solid material extending beyond $\sim$2~au and
that started with a population consisting of both km-scale
planetesimals and Moon-to-Mars mass embryos generally fail to capture
either of these two constraints (see \citealt{Raymond:2009p11530} or a
review by \citealt{Morbidelli:2012p11505}).

Recent works have explored conditions with non-smooth surface density
profiles of solids with some successes being found for a truncated
disk with an outer edge at or near 1~au
\citep{Hansen:2009p8802,Walsh:2011p12463,Izidoro:2014p15200,Jacobson:2014p18340,Levison:2015p20168}. The
altered surface density was found to promote a scattering of
Mars-analogs where they then avoid further embryo-embryo impacts and
accretion events --- essentially starving Mars, keeping it small and
ending its accretion much earlier than the Earth.

However, most previous works begin modeling at an intermediate stage of growth
- with ``planetary embryos'' amidst a sea of ``planetesimals'' - and then
truncate the disk or remove mass in certain regions.  However, this
does not consider the context of how such initial conditions came to
be. For example, if solid material never formed beyond 1.0~au, how
differently would the initial generation of planetesimals behave as
they grew into the planetary embryos?  Would there have been diffusion
of material off the sharp edges of the mass distribution? We test an
end-member case, essentially an initial disk truncation, and test if a
disk of planetesimals (diameters of 10s of km), situated between
0.7--1.0~au grow to become good matches for the terrestrial planets.

\subsection{Previous Work}

\cite{Hansen:2009p8802} explored a scenario where all solid mass
currently in the inner solar system (2~$M_\oplus$) was initially
between 0.7--1.0~au, and all of it in 400 similar sized objects (each
were 2.98$\times$10$^{22}$~kg, or $\sim$1000~km radius) with no gas
effects. These conditions succeeded in producing a good Earth/Mars
mass ratio consistently as embryos were scattered off the edge of the
annulus early and avoided further growth near 1.5~au. The accretion
timescales for Mars-analogs were fast, largely on order of the
estimated timescales from cosmochemical studies between 2-10~Myr
\citep{Nimmo:2007p11241,Dauphas:2011p19768}. However, the
Earth-analogs also formed rapidly, on roughly similar timescales,
which is much faster than cosmochemical expectations of 30-100~Myr
\citep{Kleine:2004p19341,Kleine:2009p9784}.

Seeking a mechanism to truncate a disk of solid material,
\citet{Walsh:2011p12463} invoked the inward-then-outward migration of
the Jupiter (where the \citet{Walsh:2011p12463} migration of Jupiter
is referred to as the ``Grand Tack''). This migration was constrained
by the need to produce a disk with a truncation at 1.0~au, similar to
that of \citet{Hansen:2009p8802}.  The initial conditions included 1/4
to 1/2 Mars mass planetary embryos in a sea of thousands of
planetesimals. While these initial conditions were similar to
\citet{Obrien:2006p8571} the dynamical effects of Jupiter migrating
was a powerful effect and pushed embryos onto crossing orbits. The
final planetary systems in this work had a similar mass-semimajor axis
distribution as found in \citet{Hansen:2009p8802} and later work by
\citet{Obrien:2014p13867} found similar accretion timescales.

\citet{Jacobson:2014p18340} investigated the outcomes for ``Grand
Tack'' scenarios, where the bi-modal mass distribution was altered
. Here, correlations were found such that
increasing the mass ratio of embryos to planetesimals leads to longer
accretion timescales, but also more dynamically excited final systems
of planets. Largely, the radial mass distribution (RMC) of the final planets 
was unchanged for the wide range of parameters explored.

Nebular effects have also been proposed as means to change the surface
density profile of the gas disk. Ionization in the gas-disk could lead
to regions with very different viscosity creating local mass
distribution minimums at the boundaries
\citep{Jin:2008p20003}. \citet{Izidoro:2014p15200} used this as
motivation to study planet formation in numerous scenarios with
depletions of solid material in annular regions beyond 1~au. For deep
depletions of material similar edge effects were found as in
\citet{Hansen:2009p8802} and \citet{Walsh:2011p12463}.

Finally, \citet{Levison:2015p20168} explored the growth of planets
directly from cm-sized ``pebbles''. While this work was not designed
explicitly to generate an annulus, the accretion efficiency for
pebbles (where ``pebbles'' refer to the direct accretion of cm or
small particles) is strongly dependent on both the size of the seed
body (embryo or planetesimal) and also the stopping time of the
rapidly drifting pebbles. Thus \citet{Levison:2015p20168} finds for
some initial conditions the rapid growth of embryos inside of
$\sim$1.0~au with minimal growth at further distances in the inner
solar system -- essentially generating an annulus of very large
planetary embryos.

While these works have studied formation from an Annulus in different
ways they all started modeling at an intermediate stage of growth
(note that \citet{Levison:2015p20168} relied on an entirely different
mode of accretion).  The most numerically tractable point for
commencing a model is after the bi-modal mass distribution is
established during ``Oligarchic growth'', where after a few to ten
million years there may no longer be a gas disk, and there are only
tens of embryos amidst a sea of planetesimals.  To allow a very large
number of planetesimals (thousands), the models typically do not
consider gravitational interactions between planetesimals, but the
embryos (tens of bodies typically) interact with each other and the
planetesimals. As discussed above there are variations on these
initial conditions, but nearly all are founded in this initial
bi-modal distribution of mass.

Similarly, most studies assume that when two bodies collide they merge
perfectly, conserving momentum and mass. This is caused by the
numerical problem of introducing new particles into the simulation by
way of a collisional fragmentation event and the subsequent
computational risk of large, and increasing, $N$. While various works
have included aspects of fragmentation at different times
\cite[see][]{Leinhardt:2009p10318,Kokubo:2010p9520,Chambers:2013p19990,Carter:2015p19516},
only a few models have approached planetesimal to planet simulations
\citep[see][]{Kenyon:2006p11683,Morishima:2015p19487}.

Finally, most of the works starting in Oligarchic growth phases assume
the absence of any gas effects. This is partly due to the expected
lifetime of the gaseous solar nebula (2-10~Myr) being similar to the
expected times to reach the Oligarchic growth stages. While not
computationally difficult to include, the typical absence of gas
effects is also partly due to the uncertainties in precisely when and
how the gaseous solar nebula dissipated -- whether it was a slow loss
of mass, or inside-out or outside-in dispersal. However, we will show
that in the simple case of an exponential decay of the nebula, even a
very small fraction of the original solar nebula can strongly affect
the outcome of the models by stabilizing the system for long periods
of time.

\subsection{This work}

Most previous works' initial conditions include planetary embryos, and
thus imply significant previous growth and evolution {\it before} the
various truncation/depletion mechanisms happen. None start with an
annulus of planetesimals, nor do any constrain how early or late in
the stages of growth of the disk that such truncation mechanisms could
successfully operate (technically, \citet{Levison:2015p20168} starts
with planetesimals, but studies a very different growth process).

Here, the question we are trying to answer requires a complete
simulation from planetesimals to planets, and thus re-thinking both
the initial conditions and also the gas effects requires
including particle fragmentation. Specifically, regarding a disk
truncation or annulus as a way to address the ``small Mars'' problem,
starting with a bi-modal mass distribution implies previous growth ---
the embryos have already grown. However, this would assume simple
static growth within an annulus not allowing for diffusion of bodies
or drift from drag forces. 

Thus, this test is simple, but relevant to the concept of forming the
Earth/Mars mass ratio due to substantial changes to the mass
distribution of the solid material in the disk, and may point to
required timing or truncation mechanisms to satisfy constraints.  We
will include fragmentation throughout the simulations and will include
a constantly decaying gas disk and all of its effects on the
simulation. Note that this is the first in a series of papers
exploring the growth of planetesimals to planets, and so related
studies will follow.

\section{Methods}

To model the accretion from planetesimals to planets we use the code
{\tt LIPAD}, which stands for Lagrangian Integrator for Planetary
Accretion and Dynamics \citep{Levison:2012p12338}. This code models
the fragmentation and accretion of a suite of particles, while also
modeling the dynamics of the system. This is necessary for this work
as it allows for the wholesale redistribution of mass throughout the
system unlike many collisional codes.

We describe two suites of simulations. In the first the primary
variable is the initial planetesimal size, with some tests without
fragmentation and a set of simulations with no gas effects.  These are
all refered to as the ``{\tt LIPAD} suite'' of simulations, which are
distinct from a set of ``Hansen'' simulations that re-created the
initial conditions and simulation techniques of the Hansen (2009)
work. These also tested the impact of gas effects and the results are
described in the Conclusion.

\subsection{LIPAD: planetesimals to planets}

 {\tt LIPAD} is
built on top of the $N$-body algorithms known as the Wisdom-Holman
Mappings (WHM; \citealt{Wisdom:1991p456}), and treats close
encounters between bodies using the algorithms of {\tt SyMBA}
\citep{Duncan:1998p7713}. {\tt LIPAD} utilizes ``tracer'' particles
that each represent a significant total mass for calculating the
gravitational evolution of the system.  Each tracer is given a radius
$s$, so that during the simulation it represents a swarm of particles
each of radius $s$, where the sum of their mass equals the mass of the
tracer itself.  Collisional probabilites are calculated for each
tracer as a function of its size $s$ and the total mass, sizes and
orbits of its neighbors. The dynamics of each tracer is modeled with
direct gravity calculations as well as other dynamical effects
(dynamical friction, viscous stirring), where many calculations depend
on the particle's radius $s$ and the masses, sizes and orbits of its
neighbors.

When a collision  between tracers occurs the Benz \& Asphaug
(1999) fragmentation law is used to determine the outcome. This
determines the expected size distribution of fragments, and a radius
$s$ is chosen from the distribution for each tracer involved. To
represent the full size distribution of the system requires the
inclusion of many tracer particles, each with different sizes $s$.
This algorithm has been found to match standard collisional evolution
codes due to the high number of collisions and using a high enough
particle resolution ($\sim$ thousand particles per au) so that
collisional environments can vary regionally (see Levison et
al. 2012).

When particles break, there is a minimum size that is allowable from
the collisional cascade - {\tt rfmin}. This lower limit is important
for computation times, as smaller sizes mean more collisions to
calculate. However, the physics of very small particles can be
important in the evolution of a planetary system as aerodynamic drag
drift rates increase with decreasing size down to meter-sizes and
  radiation effects start to increase below cm-sizes. Here, the
smallest tracer size allowed was 1~km. Tracers below that
size remain in the simulation but stop interacting with other tracer
particles in the collisional cascade, but still interact with larger
planetary embryos with the possibility of being accreted or ejected
from the system due to gravitational interactions.

 At the small end of the size distribution, tracers below 1~km in
size are treated as ``dust'', whereby they experience Poynting
Robertson drag with a magnitude for a 30 micron particle. This
simplification saves computational time, as resolving the size
distribution to micron, or even meter, sizes is numerically
expensive. We tested the sensitivity of this parameter, the smallest
allowable size, for a case with and without gas, where tracers could
be as small as 1~m. We found more mass loss due to collisional
grinding that was roughly $\sim$10\% greater (turning to ``dust'' and
being removed by PR drag), for the simulation setting with the 1~m
smallest tracer size. The size frequency distributions were not
substantially altered. Combined, the systematic, and relatively small,
extra mass loss, combined with minimal changes in size distributions
support our computationally expedient choice of 1~km smallest
size. This difference has been found to be much more important in
full-disk simulations (0.7--3.0 au) where the production of ``dust''
can happen in very different regions of the disk at very different
timescales, and that work will be presented in an upcoming
publication.

At the large end of the size distribution, tracers are promoted
  to embryos at $R=874$~km, at which point their mass can increase and
  their collisional evolution changes. They now revert to ``perfect
  merging'' when colliding with other embryos, while collisions with
  tracers can modify the tracer properties as described above, or
  result in a merger with the embryo.

%

Numerous gas affects are included in each simulation (see
\citealt{Levison:2012p12338} for discussions). There is a gas disk
with density 1.4$\times$10$^{-9}$~g~cm$^{-3}$ at 1~au.  The disk
density decays from this value with a 2~Myr e-folding lifetime. While
present, the gas imposes aerodynamic drag on the particles as a
function of their radius $s$, and also Type-I eccentricity damping,
which acts exclusively on the planetary embryos (see
\citealt{Levison:2012p12338} for formulations). The timescale and
  decay profile of the gaseous nebula can have profound effects on the
  outcome of these models - for the sake of simplicity while we
  explore these basic questions with a new code we will use either no
  gas or this singular decay timescale in this work.

\subsection{The {\tt LIPAD} suite}

The simulations all utilize 2612 tracer particles with randomly
generated semimajor axes between 0.7--1.0~au that follow a $1/r$ surface
density distribution. The primary variable examined was the average
initial size of the tracer particles. Three values were tested in
different subsets of simulations, where in each the particle initial
radius distribution was a gaussian centered on $R_\mathrm{init}=$3~km,
30~km and 300~km, with dispersions equal to 10\% in each case.

Particles had a density of 3~g~cm$^{-3}$ throughout, and each tracer
represented a mass of 9.35$\times10^{-4}$ M$_\oplus$ for a total of
2.44~M$_\oplus$ in the system. As growth occurs tracer particles
transition to sub-embryos when they reach a radius, $\sim$760~km and
they become full-embryos at a radius of $\sim$3400~km (see
\citealt{Levison:2012p12338} for more details on the treatment of
sub-embryos and embryos).

The initial eccentricity and inclination distributions were Rayleigh
distributions. For the $R_\mathrm{init}=$30~km and 300~km the
distribution was centered on $e=5\times10^{-5}$ and $i=0.002^{\circ}$, while
for $R_\mathrm{init}=$3~km it was centered on $e=5\times10^{-6}$ and
$i=0.0002^{\circ}$. For each initial planetesimal size, four simulations were
run for each case with different randomly generated initial
conditions. The timestep was 0.025~yr ($\sim$9~days), and the
simulations were run for 115~Myr, which take over a month computation
time running on 4-6 processors. Particles that attained a heliocentric
distance smaller than 0.045~au were removed from the simulation, as
were those that had a distance beyond 2000~au.

\subsection{The Hansen Suite}

The simulations of \citet{Hansen:2009p8802} provide an excellent (and
fast) testbed to explore the powerful effects that the gas disk can
have the evolution of a disk of solids. The simulations contain
2~$M_\oplus$ of material split between 400 bodies on orbits randomly
situated with semimajor axes between 0.7--1.0~au.  These
$R_\mathrm{init}=$1070~km planetary embryos were 0.05~$M_\oplus$
each and were randomly distributed between 0.7--1.0~au with initial
eccentricity and inclinations as Rayleigh distributions centered on
$e=0.0004$ and $i=0.01^{\circ}$.

{\tt LIPAD} was configured to allow perfect merging only, with no
fragmentation -- essentially making it a classical SyMBA or MERCURY
simulation.  There were no gas effects in one suite of four
simulations, which essentially mimic the \citet{Hansen:2009p8802}
work. In the other set of simulations a gas disk with the properties
described above was included. This disk decay time and initial density
was the same as before: 2~Myr and 1.4$\times$10$^{-9}$ gm cm$^{-3}$ at
1~au respectively.

\section{Results}

The focus of this study are the simulations described in the LIPAD
suite above. For each initial \rinit, which varied by two orders of
magnitude, there were four separate simulations performed. There were
also tests carried out with no fragmentation and no gas, and for these the
\rinit = 30~km was used. Thus we explore a singular parameter over a
wide range of values \rinit\ and two others for a fiducial value of
\rinit, and aim to have statistics based on running each multiple
times.

\subsection{Evolution and Growth of Planets in LIPAD}

\begin{figure}[h]
\includegraphics[angle=0,width=2.6in]{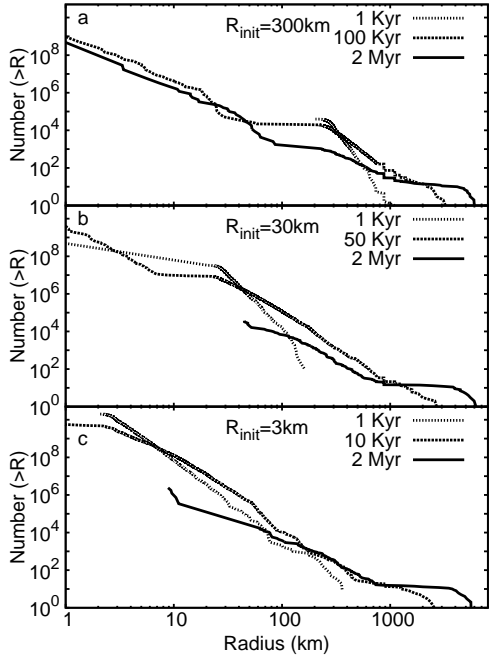}
\caption{The number of particles with radius of $N(>R)$ as a function
  of the radius of the particles $R$ for one representative simulation
  from each $R_{init}$. Pane (a) is for the simulations with
  $R_{init}=300$~km, (b) for $R_{init}=30$~km and (c) for
  $R_{init}=3$~km. Note that each pane plots slightly different
  times. The $R_{init}=3$~km reaches ``runaway'' growth much earlier
  than for the larger initial radius runs, but all three look similar
  at 2~Myr. Note, this is a not a true size frequency distribution,
  rather the representative sizes of the tracer particles in the
  simulation. }
\label{fig:SFD}
\end{figure}

\begin{figure}[h]
\includegraphics[angle=-90,width=2.8in]{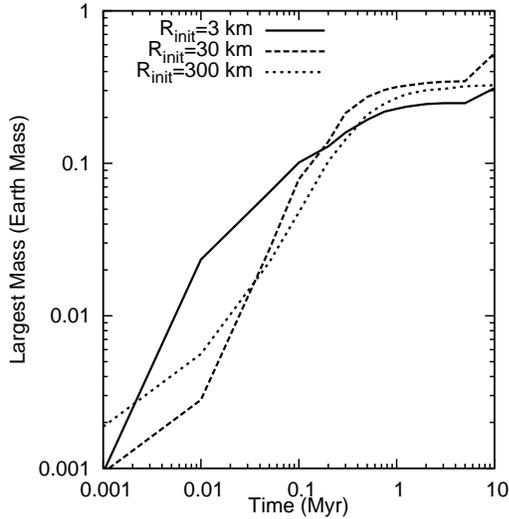}
\caption{Growth of the largest bodies in representative simulations
  for each tested \rinit, showing the mass of the largest body in
  Earth masses as a function of time.}
\label{fig:growrate}
\end{figure}

The evolution of the annulus of planetesimals followed familiar growth
patterns that have been explored extensively in the literature
progressing through two stages of growth in the first $\sim$10~Myr,
where in Fig \ref{fig:SFD} each example simulation is in the midst of
Runaway growth by 100, 50 and 10~Kyr for \rinit=300, 30 and 3~km
respectively. The annulus quickly entered Runaway growth, where
  the mass ratio between the largest body and the small bodies is
  increasing with time (see Fig.~\ref{fig:growrate}).

For each \rinit\ the growth of the largest bodies suggests that they
have reached Oligarchic growth by $\sim$1~Myr (see mass of the largest
body over time in Fig.~\ref{fig:growrate}), where Oligarchic growth is
typically considered to be begin when the stirring power of the large
bodies exceeds that of the small bodies
\citep{1993Icar..106..210I}. Therefore the size at which this
transition happens has a functional dependence on the initial
planetesimal size - the size of the small bodies. In terms of radius,
for the three sets of intial conditions \rinit=300, 30 and 3~km this
transition should take place when bodies grow to 440~km, 110~km and
28~km \citep{Ormel:2010p11708,2003Icar..161..431T}.  Oligarchic growth
continues until embryo-embryo collisions begin and the onset of
instabilities begins closer to $\sim$10~Myr (see Sec.~\ref{instable}
below for characterization of instabilities and the onset of the Giant
Impact stage).

\begin{figure}[h]
\includegraphics[angle=-90,width=2.5in]{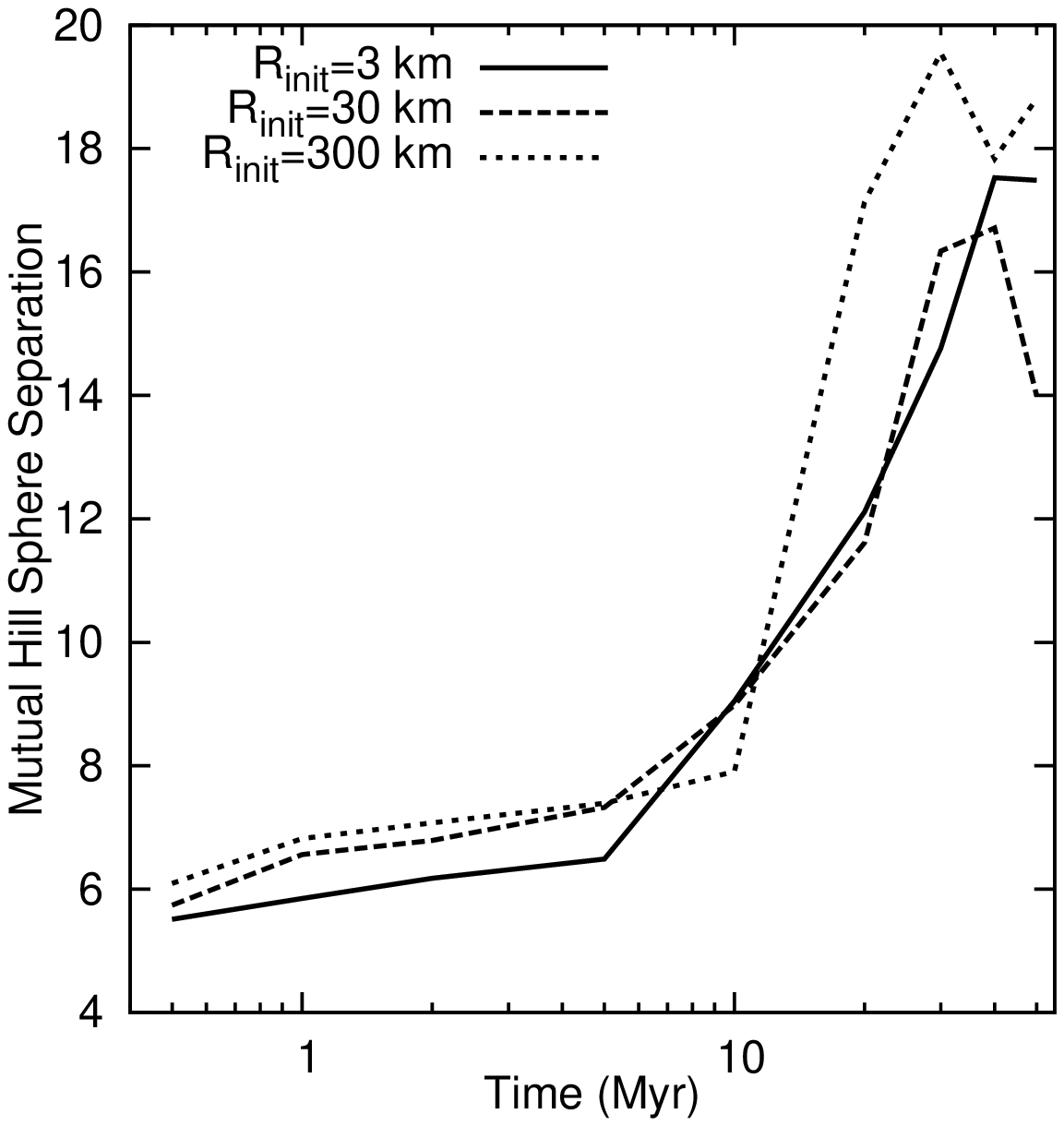}
\caption{Embryos spacings in units of mutual Hill Spheres, for all of
  the simulations with \rinit = 3, 30 and 300~km, as a function of
  time.}
\label{fig:HillTime}
\end{figure}

The spacing of the planetary embryos (where anything larger than
1/70th the mass of the Earth is included for this calculation as an
``embryo'') is similar for each set of initial conditions
(Fig.~\ref{fig:HillTime}). At 1~Myr, where the simulations had on
average 12 embryos each, the distribution of spacing in terms of
mutual Hill Spheres peaks around 6 for each \rinit. The spacing, in
terms of semimajor axis, slowly increases with time for all cases,
until $\sim$10~Myr, at which time the spacing starts to grow
substantially as simulations start to enter the Giant Impact stage of
growth. This evolution is in line with the expectations and previous
results for the spacing of embryos during the Oligarchic growth stage
\citep{Kokubo:1998p9706}.

\begin{figure}[h]
\includegraphics[angle=-90,width=2.5in]{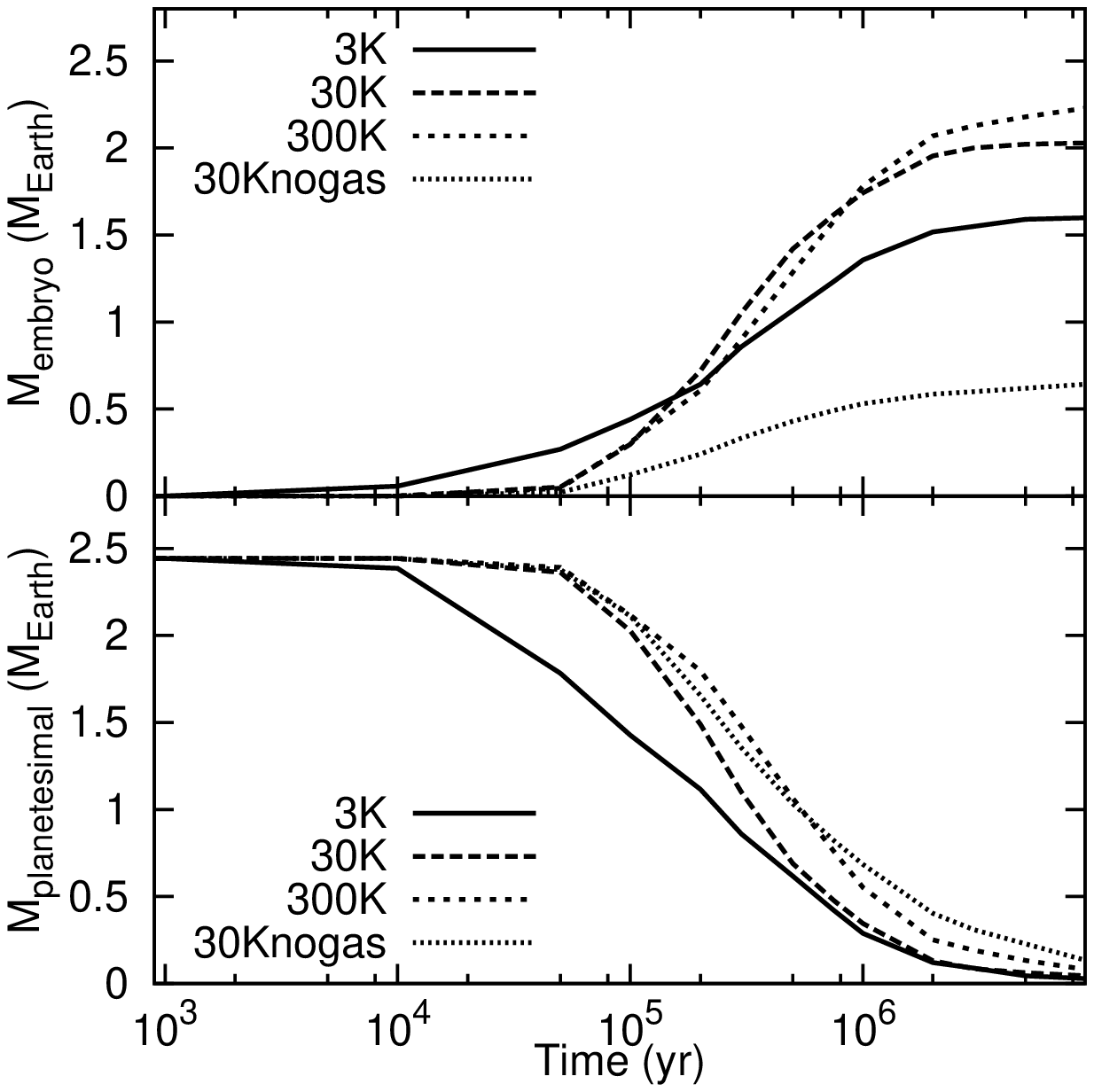}
\caption{The (upper panel) total mass of embryos in units of Earth
  Masses as a function of time for three nominal simulations at each
  \rinit, and also for the \rinit = 30~km case with no gas. The
  (bottom panel) total planetesimal mass as a function of time for the
  same simulations.}
\label{fig:MassEvol}
\end{figure}

Mass can be lost from the simulations due to the inclusion of
fragmentation, gas drag effects and Poynting Robertson (PR) drag.  In
Figure \ref{fig:MassEvol} a clear trend for more surviving final mass
in embryos is found for increasing \rinit\ with only 1.6, 2.0 and
2.2~$M_\oplus$ remaining for 3, 30 and 300~km respectively (where the
embryo mass is a proxy for system mass as planetesimal mass approaches
zero in all cases). This is due to more efficient grinding at smaller
sizes. The smallest \rinit\ shows the fastest growth in the first
100,000 years and the earliest production of embryos, which would stir
the neighboring disk and promote collisional grinding earlier. The case with
no gas disk shows substantial mass loss, with over half of the mass
lost from the system, due to the dynamical excitement early in the
disk evolution when there were still many small planetesimals to grind
away.

\subsection{The Planets}

In this work we consider ``Planets'' to be anything over 1/30th the
mass of the Earth (unlike many classical works we grow our embryos and
so we selected $\sim$1/2 Mercury mass to cover the entire range of
planetary mass). We define a Mars-analog as any planet that has a
semimajor axis between 1.2 and 2.0~au. Therefore some simulations may
have multiple Mars-analogs (though this is rare in these simulations -
and we attempt to indicate the cases and discuss further
individually).

\subsubsection{Bulk properties}
The properties of the formed planetary systems are listed in Table
\ref{bigtable}. The average number of planets formed for each suite of
4 simulations was 3.5, 3.25 and 5.25 for \rinit = 3, 30 and 300~km
respectively, and 3.75 for \rinit = 30~km with no gas. The total mass
remaining in the planetary systems decreases with \rinit, averaging
over 2~$M_\oplus$ for \rinit = 300~km, and 1.26 $M_\oplus$ for \rinit
= 3~km. This is due the affects of collisional grinding as mentioned
above. For the case with no gas the grinding was more severe,
and the mass loss was due to grinding to very small sizes and drag 
removing particles --- these cases ended with below 1~$M_\oplus$ of
planets.

One can rely on some typical metrics to assess the mass distribution
and orbital excitement compared to the current system of Terrestrial
Planets and previous modeling efforts. To measure the mass
distribution, we use the radial mass concentration statistic (RMC),
defined as

\begin{equation}
RMC = max\bigg(\frac{\sum M_j}{\sum M_j[\log_{10}(a/a_j)]^2}\bigg) ,
\end{equation}

\noindent where $M_j$ and $a_j$ are the mass and semimajor axis of
planet $j$ \citep{Chambers:2001p7618}. The function in brackets is
calculated for a range of $a$ (0.7--1.2~au) in the terrestrial planet
region with the maximum value being selected (following the
prescription of \citealt{Chambers:2001p7618}). This metric returns
infinity for a single planet system, and decreases towards zero as
mass gets spread out across a range of semimajor axes.  The current
value for the solar system is 89.9. This value exceeds all outcomes
for \rinit = 30 and 300~km, where their low values point to a wider
distribution of masses (see Table \ref{bigtable}). For example for
\rinit=300~km, run (003), the RMC is skewed low due to the presence of
a large (0.57~$M\oplus$) and distant ($a=1.7$~au) Mars-analog. The
simulations for \rinit=3~km and the \rinit=30~km with no gas typically
exceeded the solar system RMC -- sometimes due to a small number of
centrally (near 1~au) located planets (as in one case for \rinit=3~km
that produced only 2 planets).

A common metric used to measure the dynamical excitation of a
planetary system is the angular momentum deficit (AMD; Laskar 1997):

\begin{equation}
AMD = \frac{\sum_j M_j \sqrt{a_j}\left(1-\cos(i_j)\sqrt{1-e_j^2}\right)}{\sum_j M_j \sqrt{a_j}}  ,
\end{equation}

\noindent where $M_j$ and $a_j$ are again the mass and semimajor axis,
while $i_j$ and $e_j$ are the inclination and eccentricity of each
planet $j$. The current solar system has an AMD = 0.0014. All of the
\rinit=3 and 30~km simulations had final AMD values exceeding the
solar system value, typically by a factor of 2--10. The two
\rinit=300~km cases with similar or lower AMD values were the two
cases with 7 planets. Generally, as will be discussed below, the
systems had depleted nearly all planetesimals before having any major
orbital instability, leaving behind insufficient masses of
planetesimals to significantly damp the orbits of the final
planets. Note, that as will be discussed in the Conclusions below, the
way that this code handles the fragmentation of embryos may change the
amount of damping provided by planetesimals produced {\it during} the
final few giant impacts.

\begin{figure}[h]
\includegraphics[angle=-90,width=2.5in]{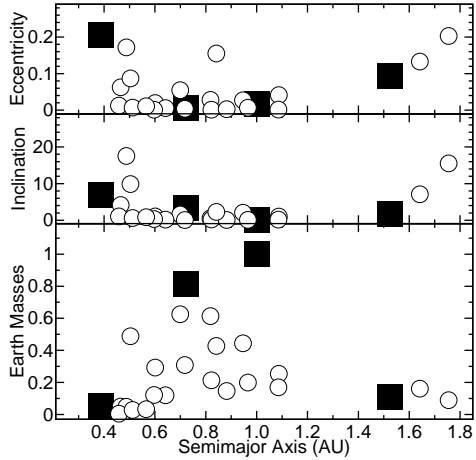}
\caption{The final planets, after 115Myr of evolution, for \rinit =
  3~km. Plotted are the (top) eccentricity, (middle) inclination in degrees and
  (bottom) mass as a function of semimajor axis. The large black
  squares are the values for Mercury, Venus, Earth and Mars, while the
  open circles are the simulation outcomes.}
\label{fig:Planets3k}
\end{figure}

\begin{figure}[h]
\includegraphics[angle=-90,width=2.5in]{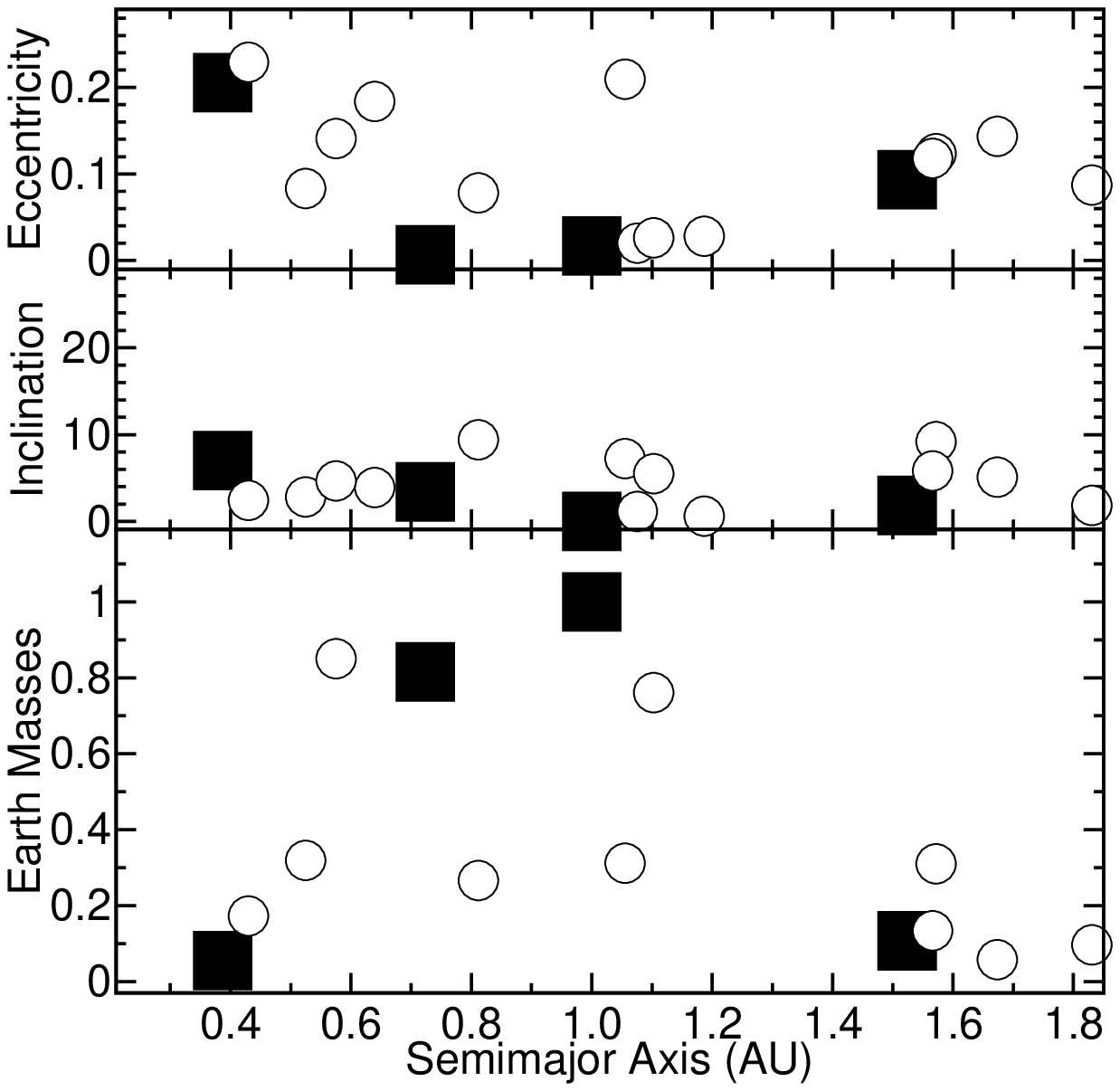}
\caption{The final planets, after 115Myr of evolution, for \rinit =
  30~km. Plotted are the (top) eccentricity, (middle) inclination in degrees and
  (bottom) mass as a function of semimajor axis. The large black
  squares are the values for Mercury, Venus, Earth and Mars, while the
  open circles are the simulation outcomes.}
\label{fig:Planets30k}
\end{figure}

\begin{figure}[h]
\includegraphics[angle=-90,width=2.5in]{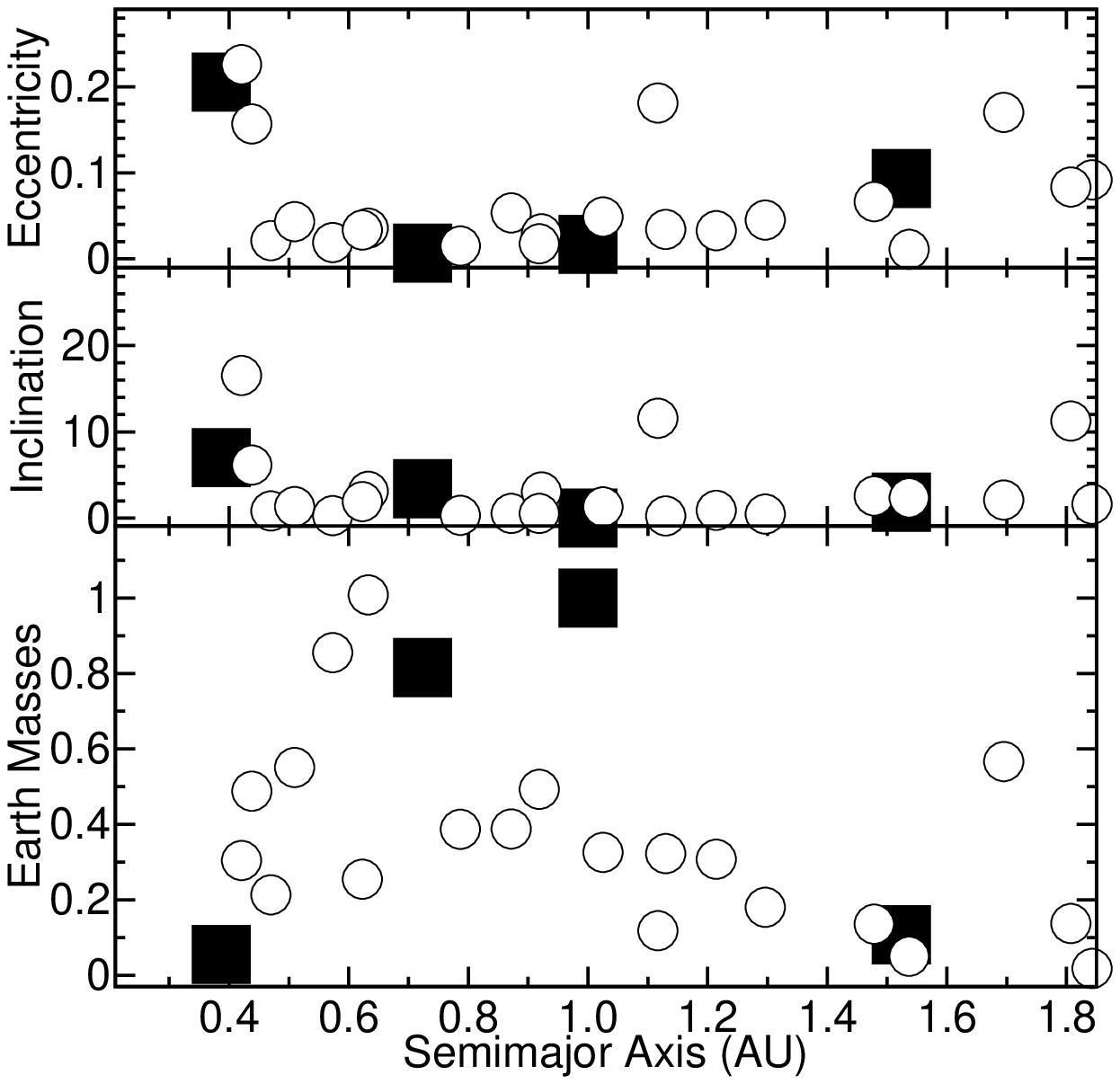}
\caption{The final planets, after 115Myr of evolution, for \rinit =
  300~km. Plotted are the (top) eccentricity, (middle) inclination in degrees and
  (bottom) mass as a function of semimajor axis. The large black
  squares are the values for Mercury, Venus, Earth and Mars, while the
  open circles are the simulation outcomes.}
\label{fig:Planets300k}
\end{figure}

Ten of the twelve baseline simulations (all the cases with gas
described so far) produced a Mars-analog, and 13 total Mars-analogs
were produced - where a Mars-analog is any embryo between 1.2 and
2.0~au (\rinit=300 cases (001) and (004) each had two Mars-analogs,
along with \rinit=3~km (003)). The Mars-analogs were on average 0.16
M$_{\oplus}$ and located at 1.56~au - these simulations
produce reasonable Mars-analogs. The low RMC values values discussed
above are therefore not due to Mars-analogs being large, but rather
due to the depressed masses of, or absence of, Earth- and
Venus-analogs. This is notable in the figures of each set of
simulations mass compared with the semimajor axis, where planets at
1.0~au are often only half the mass of the Earth (see Figures
\ref{fig:Planets3k}, \ref{fig:Planets30k}, \ref{fig:Planets300k}).

\subsection{Instabilities and Last Giant Impacts}\label{instable}

Many of the simulations showed long periods of quiscient behavior,
followed by a short interval of crossing orbits, embryo-embryo impacts
and a rapid decrease in the number of embryos or planets. These
instabilities were tracked using the metric of recording the beginning
of a 1~Myr interval where at least two embryo-embryo collisions
occured. While inexact, this metric provides a quick look at when a
system evolves from a dynamically cold suite of many planets and
reduces its numbers (this metric, $t_\mathrm{inst}$ is reported in the
table of results Table \ref{bigtable}).

The similarity in instability times was somewhat surprising, where,
for example in the \rinit=300~km cases, all four simulations had at
least one instability between 10.5--13.7~Myr. The 12 baseline test
cases all had instabilities between 8.4--19.8~Myr. There is no
apparent connection between any of the studied outcomes (number of
planets, Mars-analogs etc.) and the existence or timing of the
instabilities.

The last giant impacts, t$_\mathrm{Limp}$, were determined by last
impact that included two planetary embryos. The timing of these were
not clearly correlated with \rinit, or the other instability
times. Each suite of \rinit\ had some relatively early last impacts,
$<$25~Myr, and some relatively late impacts $>$75~Myr.

\begin{figure}[h]
\includegraphics[angle=-90,width=2.5in]{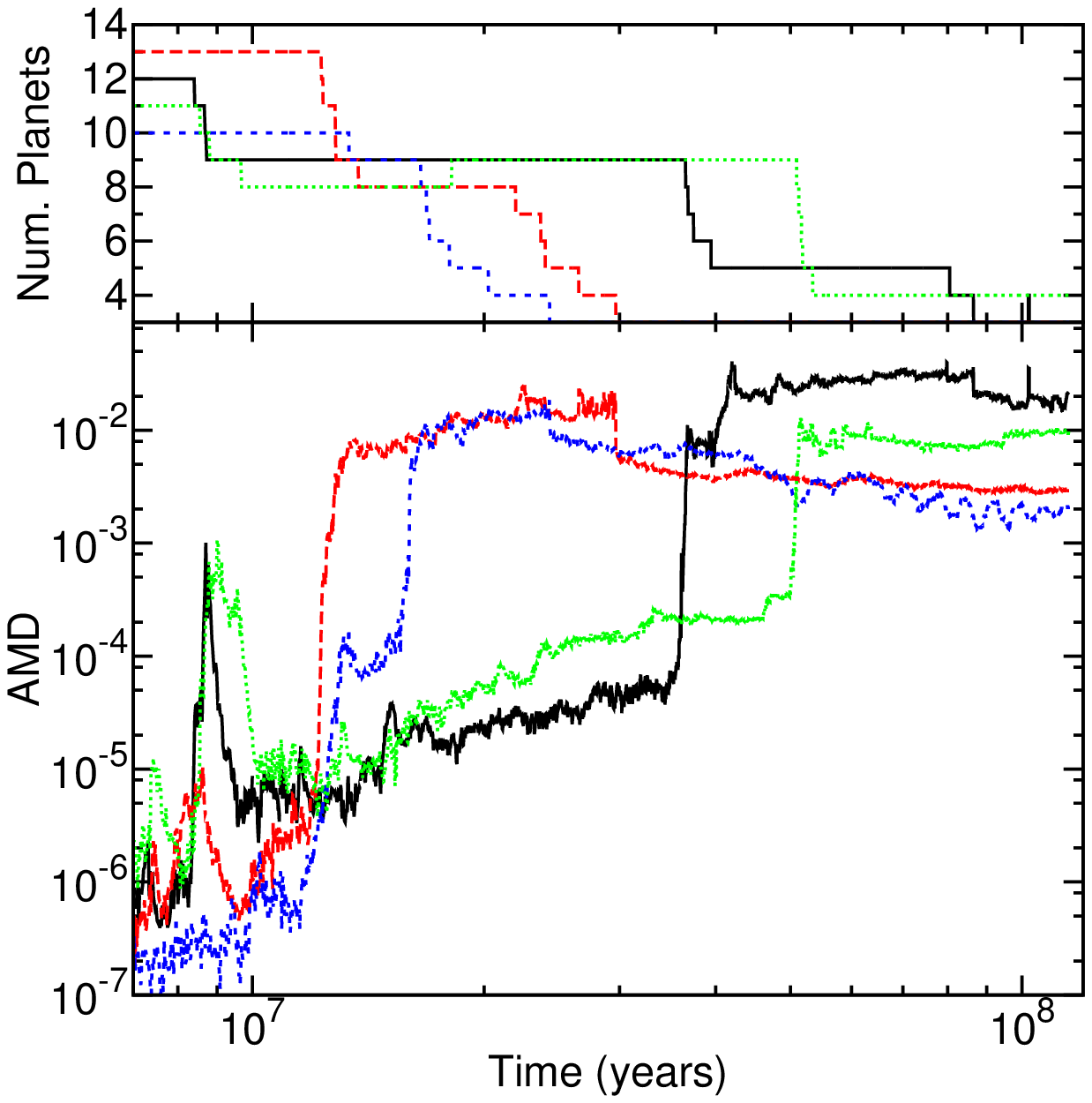}
\caption{The (top) decrease in number of planets in simulations
  correlated with the (bottom) calculated AMD values as a function of
  time. This shows all four simulations for the
  R$_\mathrm{init}$=30~km case.}
\label{fig:AMDevol}
\end{figure}

The clearest correlation is the rise of AMD values with the late
decrease of number of planets in each simulation. Plotted in
Fig. \ref{fig:AMDevol} is the evolution of the 4 different \rinit =
30~km cases evolution of number of planets and the corresponding AMD
values as a function of time. The AMD values remain relatively low
beyond 10~Myr (with occasional spikes), and only rise to their final
elevated levels at the beginning of the final instability (as shown by
the drop in $N_\mathrm{pl}$ in the top pane). The once dynamically cold
systems, with low AMD, get dynamically excited, and start losing
planets to giant impacts, until their final planetary suite is
reached.

The instability times found here, range from 8~Myr to 18~Myr,
  which for the gas disk dissipating with a 2~Myr e-folding time,
  finds that between 1.8\% and 0.12\% of the initial gas density is
  present. Previous works focusing on orbital stability of systems as
  a function of their Hill spacing and gas disk surface density find
  that less than 0.1\% of the nebular gas (in terms of MMSN) must
  remain to find instabilities for systems with typical Hill spacing
  ($\sim$10 Hill radii)
  \citep{2001DPS....33.1507I,2002Icar..157...43K}. Our findings are in
  line with this previous work and together point to an important
  dependence on a system's initial instability on a very small
  fraction of an initial gaseous nebula.

Another way of interpreting this behavior is that there are, in
affect, two generations of planets. The first is generated from
Oligarchic growth, where the embryos reach a dynamically stable
configuration due to continued damping effects of the gas and also due
to the ability for the annulus to diffuse outward. Then the passage
into the Giant Impact phase is sudden and only begins 10's Myr later.

\begin{table*}[p]
\resizebox{\linewidth}{!}{%
\begin{tabular}[h]{l|ccccccccc}
 R$_\mathrm{init}$ (run \#) & N$_\mathrm{planets}$ & M$_\mathrm{tot}$ & AMD  & RMC & M$_\mathrm{Larg}$ (M$_\oplus$) & M$_\mathrm{M}$ (M$_\oplus$)&a$_\mathrm{M}$ (au) & t$_\mathrm{Limp}$& t$_\mathrm{inst}$\\
\hline 
3 km  (001)  & 4 & 1.43  & 0.00252    & 61.8  & 0.74 &0.11 & 1.47 &34.1 &11.0\\
3 km  (002)  & 4 & 1.08  & 0.00914    & 94.7  & 0.51 &    &      &20.8 &14.0,16.3\\
3 km  (003)  & 4 & 1.24  & 0.01274    & 59.1  & 0.84 &0.14& 1.27 &28.6 &14.4\\
3 km  (004)  & 2 & 1.27  & 0.00564    & 109.7 & 0.79 &    &      &75.4 &19.8\\
\hline 
Average      & 3.5& 1.26 & 0.00751    & 80.6 &  0.72 & 0.13 & 1.37 & 39.7 & 14.8 \\
\hline 
30 km  (001)  & 3  &1.82  & 0.02161    & 42.9 & 1.20 &0.31 &1.57 &86.4 &8.4,36.6\\ 
30 km  (002)  & 3  &1.57  & 0.00292    & 49.2 & 1.34 &0.06 &1.67 &29.7 &12.2,23.6 \\
30 km  (003)  & 3  &1.75  & 0.00209    & 56.2 & 1.29 &0.13 & 1.57 &24.3 &16.6\\
30 km  (004)  & 4  &1.97  & 0.00939    & 44.3 & 0.85 &0.10 & 1.83 &53.5 &8.6,50.9\\
\hline 
Average       & 3.25 & 1.78&0.00900    & 36.9 & 1.17 & 0.15& 1.66 & 48.5 & 11.5\\
\hline
300 km (001)   & 7 &2.22   & 0.00101    & 40.4 & 0.85 & 0.31 & 1.21 &13.6 & 10.5,11.9\\
300 km (002)   & 3 &2.06   & 0.00665    & 36.8 & 1.44 &0.14 & 1.80 &104.6 & 11.6,13.3\\
300 km (003)   & 4 &1.99   & 0.01643    & 20.6 & 1.00 &0.57 & 1.70 &78.8 & 11.7, 13.0\\
300 km (004)   & 7 &2.23   & 0.00065    & 49.5 & 0.55 &0.18 & 1.30 &16.8 &13.7\\
\hline 
Average        & 5.25&2.13 & 0.00619    & 36.8 & 0.96 &0.30 & 1.43 & 53.5 & 11.9\\
\hline
30 km no gas (001)  & 5  &0.67  & 0.00089    & 162.4 & 0.29 &  &        & 21.8    &20.2 \\ 
30 km no gas (002)  & 3  &0.62  & 0.00821    & 94.7 & 0.35 &  &         & 71.5    & \\
30 km no gas (003)  & 3  &0.61  & 0.00707    & 71.9 & 0.37 &0.04 & 1.55 & 101.2   & \\
30 km no gas (004)  & 4  &0.63  & 0.00168    & 120.2 & 0.26 &0.02 &1.38 & 10.9    & \\
\hline
Average             &3.75&0.63  & 0.00446    & 112.3 & 0.32 & 0.03 &1.47 & 51.4 & 20.2 \\
\hline 
30 km no Frag (001)                & 3 & 1.82 & 0.02004  & 39.1 & 1.16 &0.42 &1.24 &56.9 &18.5\\
\hline
1070 km no Frag no gas (001)       & 6 & 1.96 & 0.00740  & 66.0 & 0.88 &0.04 & 1.85 & 12.5 &1.2,4.0,8.9,10.4\\ 
1070 km no Frag no gas (002)       & 6 & 1.90 & 0.01758  & 32.7 & 0.86  &0.07 & 1.82 & 70.0 &1.3,6.1,8.1\\ 
1070 km no Frag no gas (003)       & 8 & 1.88 & 0.00899  & 45.8 & 0.94 &0.23 & 1.58 &101.3 & 1.0, 3.9\\
1070 km no Frag no gas (004)       & 6 & 1.93 & 0.00526  & 78.0 & 0.91  &0.04 & 1.66 &3.6 & 1.0,11.5\\
\hline 
Average                            & 6.5&1.91 & 0.00981  & 55.6 & 0.90 & 0.10 & 1.73& 46.9  & 1.1 \\
\hline 
1070 km no Frag with gas (001)  & 4  &1.99  & 0.04572    & 27.7 & 0.86 &0.31 & 1.82 & 104.3 &17.1\\ 
1070 km no Frag with gas (002)  & 5  &1.99  & 0.01689    & 43.2 & 0.98 &0.38 & 1.38 & 113.5 &92.8 \\ 
1070 km no Frag with gas (003)  & 4  &1.99  & 0.01231    & 41.6 & 1.03 &0.19 & 1.43 & 27.1  &13.3,17.5\\
1070 km no Frag with gas (004)  & 3  &1.87  & 0.01389    & 47.5 & 0.95 &0.53 & 1.24 & 40.8  &\\
\hline 
Average                         & 4 &1.96   & 0.0222     & 40.0 & 0.96 &0.35 & 1.47 & 71.4  & 41.1 \\
\hline 

\end{tabular}}
\caption{Total mass of planets M$_\mathrm{tot}$, and the mass of the
  largest planet in the system, M$_\mathrm{Larg}$, at 115~Myr.  Mass
  of Mars analogs (M$_\mathrm{M}$; if $1.2<a<2.0$) and their semimajor
  axis at completion (a$_\mathrm{M}$). The time of the last
  embryo-embryo impacts, t$_\mathrm{Limp}$, where an embryo is
  considered if it is 1/30 M$_\oplus$. The time of any major
  instabilities (t$_\mathrm{inst}$) that result in at least two
  planet-planet mergers within 1~Myr. Note that 3~km run 003 had a
  0.04~M$_\oplus$ at 1.62~au. \label{bigtable} }
\end{table*}

\subsection{The effect of the Gas Disk and Fragmentation} \label{NoGas}

\begin{figure}[h]
\includegraphics[angle=-90,width=2.5in]{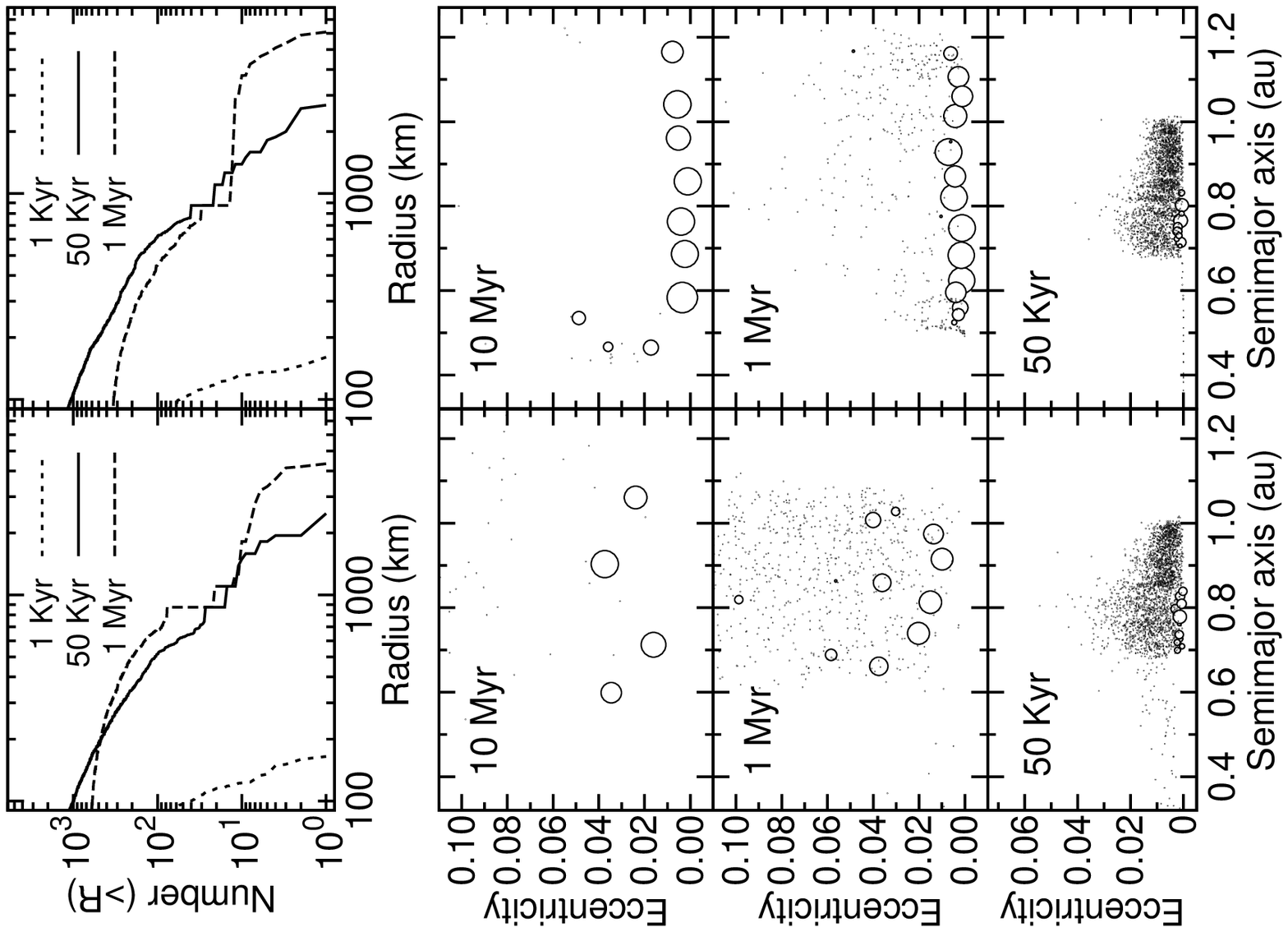}
\caption{ The top two panes show the size distribution in each of the
  simulations with no gas (left) and those with gas (right) at three
  different times. The bottom panes show the evolution of these two
  simulations in terms of eccentricity and semi-major axis (au).}
\label{fig:NoGasEvol}
\end{figure}

\begin{figure}[h]
\includegraphics[angle=-90,width=2.5in]{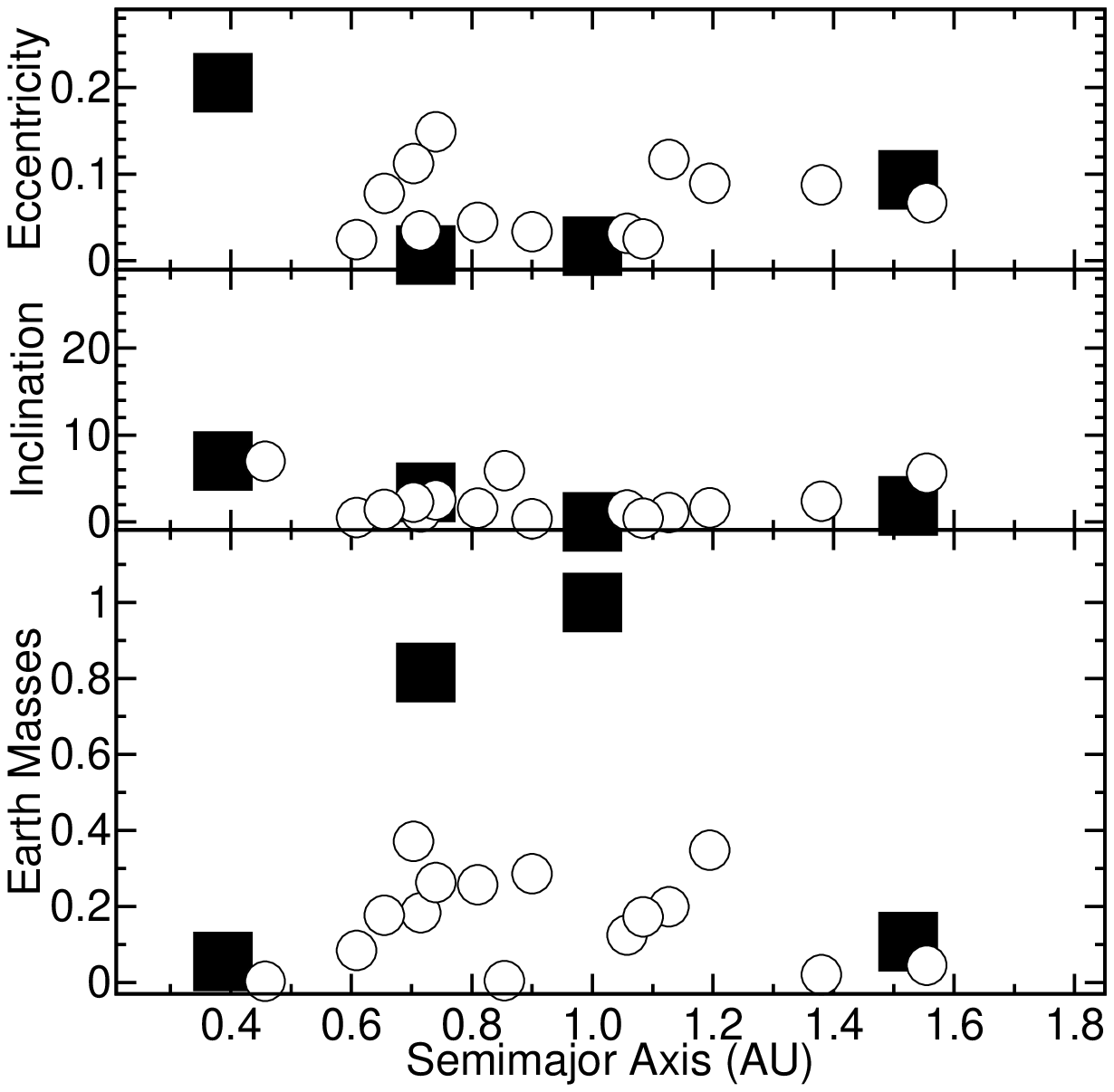}
\caption{The final planets, after 115Myr of evolution, for \rinit =
  30~km and no gas affects. Plotted are the (top) eccentricity, (middle) inclination in degrees and
  (bottom) mass as a function of semimajor axis. The large black
  squares are the values for Mercury, Venus, Earth and Mars, while the
  open circles are the simulation outcomes.}
\label{fig:Planets30kNoGas}
\end{figure}

The gas disk plays a large role in the outcome of these
simulations. The damping provides an environment for rapid growth of
embryos and diffusion in semimajor axis without crossing orbits (see
Fig. \ref{fig:NoGasEvol}). The evolution for the suite of cases run with
no gas disk is starkly different than that found for all cases tested
with the gas disk discussed above (see
Fig. \ref{fig:Planets30kNoGas}). 

\begin{figure}[h]
\includegraphics[angle=-90,width=2.5in]{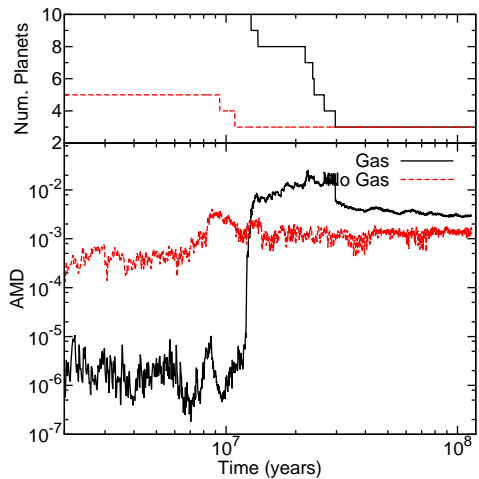}
\caption{The (top) number of planets and (bottom) AMD plotted as a
  function of time for one case with gas (solid black line) and for a
  case with no gas (dashed red line).}
\label{fig:GasNoGasAMD}
\end{figure}

However, the final systems are, in terms of number of planets and
relative distribution of remaining mass, not vastly different than the
\rinit=30~km simulations with gas. In fact, as will be discussed at
length in the Conclusion below, and explored with more test cases,
these point to the curious scenario where very different evolutionary
paths of two systems reach a similar end state. Figure
\ref{fig:GasNoGasAMD} shows the evolution for a case with gas and case
without gas in terms of their number of planets and AMD. Both reach 3
planets and have similar final AMD, but they do so on very different
timescales and with a vastly different evolution of the system AMD.

\begin{figure}[h]
\includegraphics[angle=-90,width=2.5in]{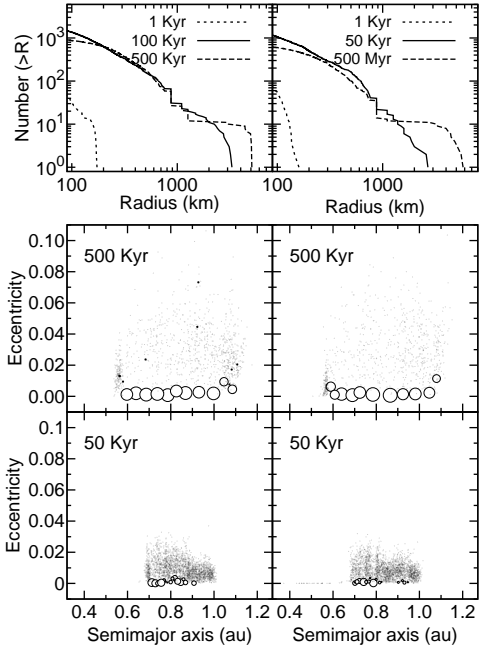}
\caption{Evolution of two simulations, the left column are two panels showing the growth and evolution of the size frequency distribution without fragmentation, while the right panels include fragmentation.}
\label{fig:NoColl}
\end{figure}

\begin{figure}[h]
\includegraphics[angle=-90,width=2.5in]{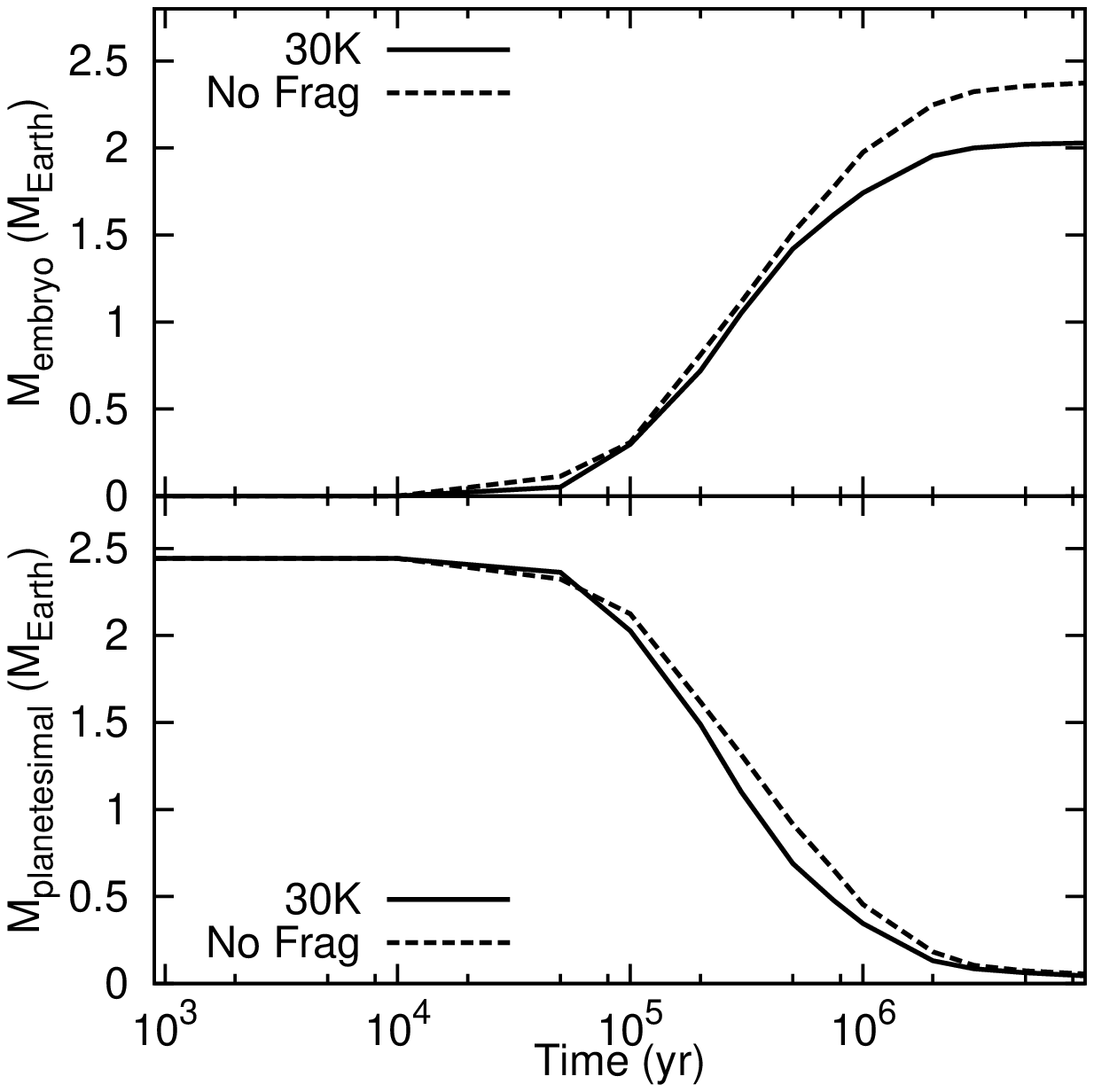}
\caption{Evolution of two simulations, the top shows the growth of embryo mass over time and the bottom the total planetesimal mass over time for the nominal simulation with \rinit = 30~km and the case with no fragmentation.}
\label{fig:NoCollMassEvol}
\end{figure}

While the effects of the gaseous solar nebula were extreme, the
effects of fragmentation were less obvious.  To test the effects of
collisional evolution on the outcome of the simulations a test was run
where particles were allowed to grow, but not fragment, during
collisions. The \rinit\ tested was 30~km, and the evolution was
qualitatively similar to that found for the model with fragmentation
(see Fig. \ref{fig:NoColl}). The clear difference was the total mass
in the system, whereby without fragmentation all of the mass stayed in
the system and was accreted by the system of embryos (see
Fig. \ref{fig:NoCollMassEvol}). For nominal ordered growth in the
presence of a long-lasting disk (Myr timescales), the effects of the
collisional cascade are not critical -- an external source of dust to
damp orbits, or a mechanism for dynamic excitation, would be needed to
drastically alter the outcomes.

\subsection{Comparison with the Hansen suite}\label{hansendisc}

Using the various metrics explored above we find that the final
planetary system outcomes from these suites of simulations produce
adequate matches of the solar system. On average the number of planets
is slightly too low and the orbits are too excited. However,
Mars-analogs are regularly reproduced, though the mass-semimajor axis
distribution (see the RMC metric for the \rinit=30 and 300~km cases)
suggests that the Earth-analogs are too small and the systems are a
bit too spread out. However, for the nominal case of \rinit=30~km the
systems are reasonable matches.

This is surprising because the evolution of the \rinit=30~km systems
is decidedly different than the evolution of the successful systems in
\citet{Hansen:2009p8802}. To highlight and explore this further we ran
a small series of simulations designed to mimic very directly (and
quickly) the original \citet{Hansen:2009p8802} simulations with and
without gas using {\tt LIPAD} but without including fragmentation of
the tracer particles (just 400 interacting planetary embryos - this is
the ``Hansen Suite'' described above).

\begin{figure}[h]
\includegraphics[angle=-90,width=2.5in]{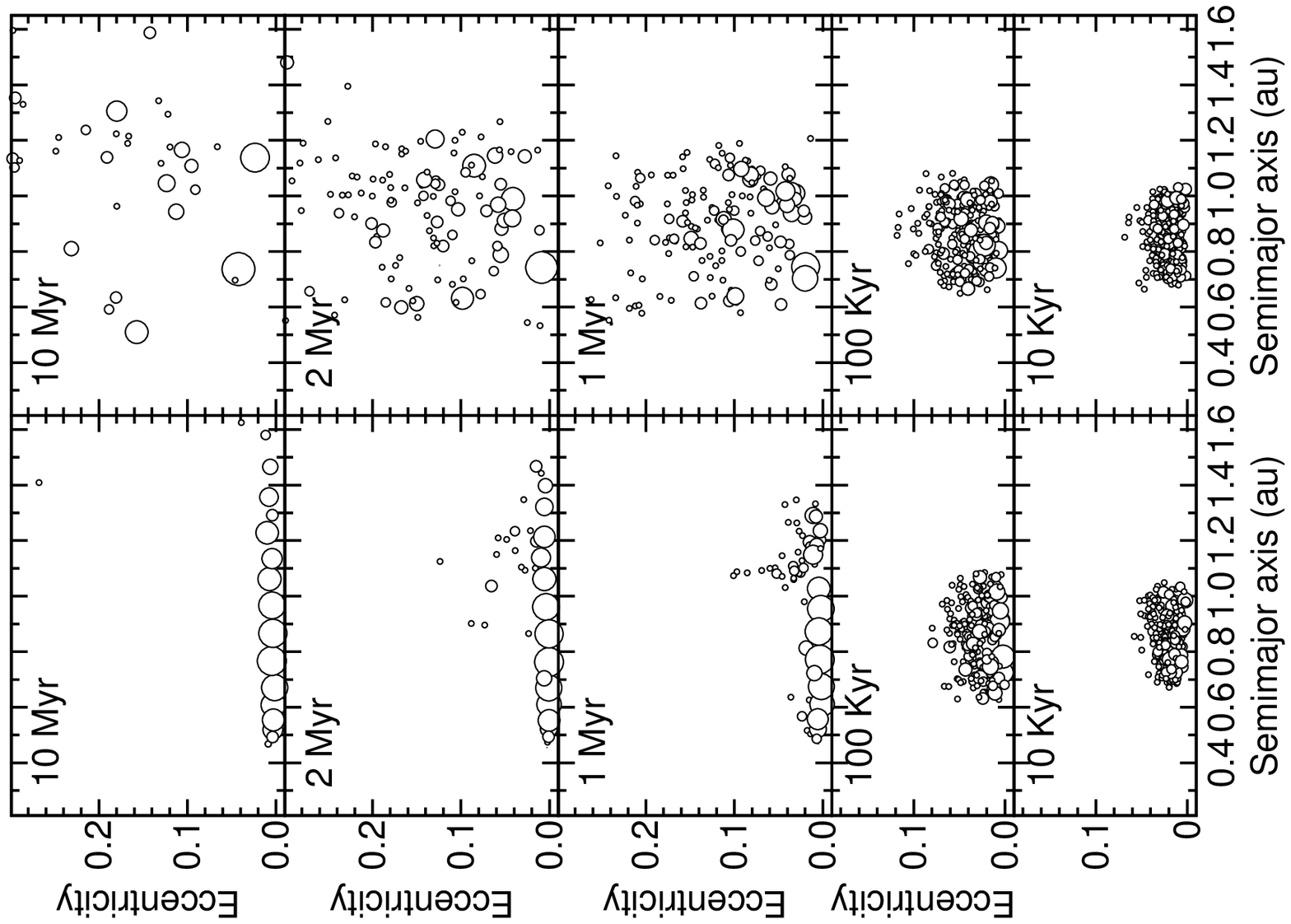}
\caption{Comparing initial conditions from Hansen (2009) with and
  without a gas disk (left and right respectively).}
\label{fig:HansenFrames}
\end{figure}

The evolution is very different (see Figure \ref{fig:HansenFrames}
left pane) for the simulation using a gas-disk (2~Myr e-folding
time). After about $\sim$1~Myr there are a small number of embryos,
well spaced, and dynamically cold. Meanwhile, for the simulations with
no gas the system is very dynamically excited by 1~Myr (see Figure
\ref{fig:HansenFrames} right pane), and the size distribution is not
bi-modal, instead having a few embryos as large as found in the
gas-disk case and a range of smaller embryos.

\begin{figure}[h]
\includegraphics[angle=-90,width=2.5in]{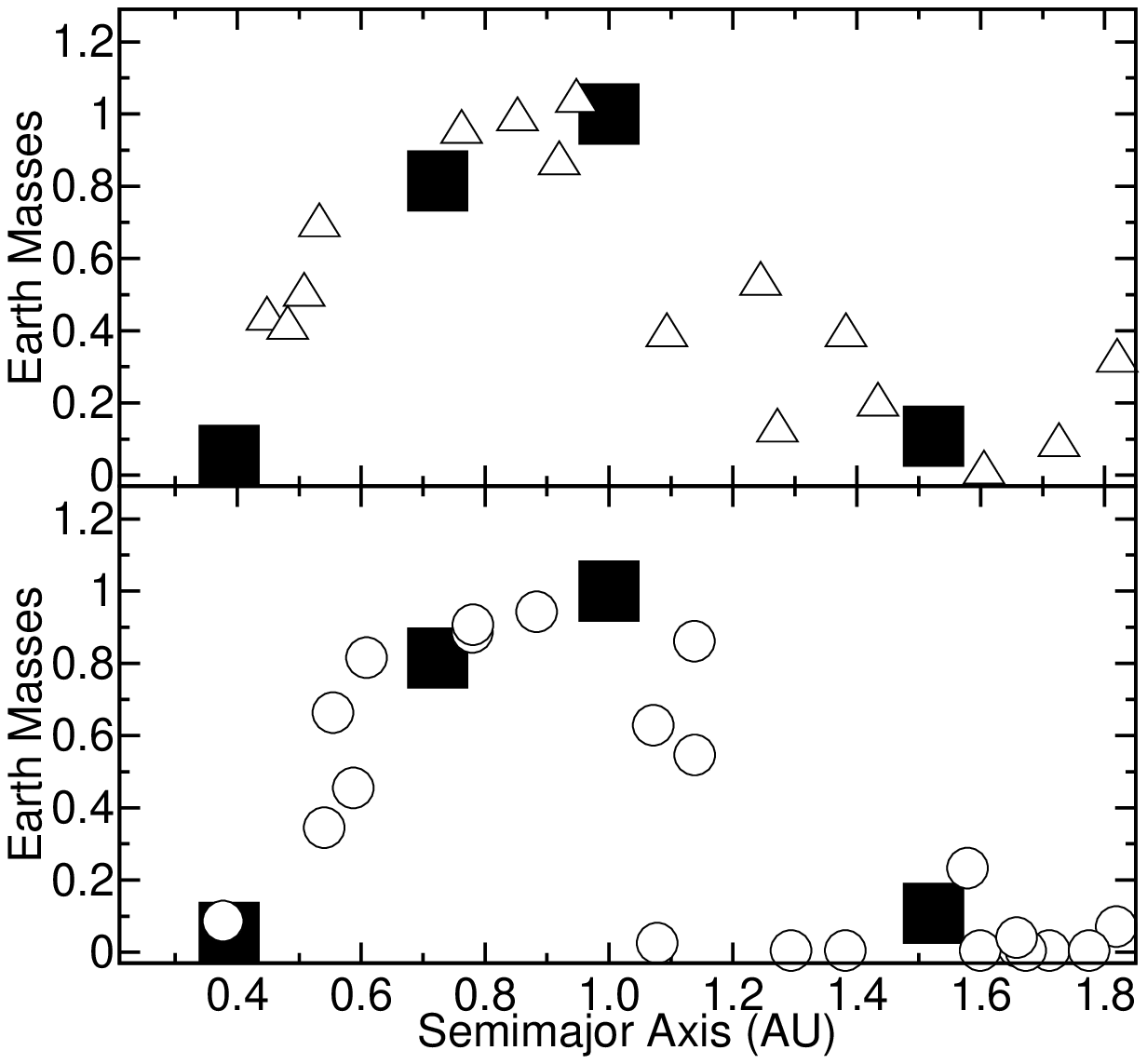}
\caption{The top pane (open triangles) shows the orbital element distribution of
  planets using the Hansen 400 embryos initial conditions simulated in
  {\tt LIPAD} with a gas-disk that dissipates on a 2-Myr timescale. The
  bottom pane (open circles) shows the same set of simulations run with no gas disk,
  to more directly reproduce the simulations of Hansen (2009).}
\label{fig:HansenPlCompare}
\end{figure}

The final systems from this test are remarkably similar to each other
(see Fig.~\ref{fig:HansenPlCompare}). The cases with the 2-Myr gas
disk evolve into an Oligarchic bi-modal distribution with nearly all
mass in embryos, and only later evolve through an instability to break
from a cold system of many planets to reach a final system
observed. Meanwhile the simulations with no gas stay excited with
regular embryo-embryo impacts and maintain significant dynamical
excitement throughout.

This divergent evolution with similar final results mirrors that seen
in the {\tt LIPAD} suite of simulation that behaved similar to these
tests run with gas effects. Despite evolving through a long quiescent
period, and having two generations of planets, the final systems are
similar.

\section{Conclusions}

The intent of this study was to explore whether there were limits on
building good Mars-analogs and Earth/Mars mass ratios from a truncated
disk or annulus. The end-member case of an annulus of km-sized
planetesimals was modeled including gas and fragmentation effects, and
found in most cases to produce adequate matches to the observed
terrestrial planets, including Mars analogs.

However, as explored above, while these may produce similar final
systems of planets as in the original \citet{Hansen:2009p8802} work
the evolution of the system was vastly different. The systems final
properties are therefore potentially less diagnostic of the physics
involved in their growth than is the expression of a planet's
accretion profile in geochemical data.  Rather, an approach that
considers the constraints on growth timescales and also the
compositional and internal evolution due to different growth profiles
may be more important in assessing all of the important factors in
Terrestrial Planet formation
\citep{Fischer:2014p20911,Dwyer:2015p20491,Carter:2015p19516}.

Generally, a Mars-analog in these simulations accumulates most of its
mass rapidly through the accretion of planetesimals, avoids late
embryo-embryo collisions and its orbit diffuses outward from
$\sim$~1~au to its final semimajor axis at $\sim$1.5~au.  This growth
profile for Mars-analogs is not too different than that found in
\citet{Hansen:2009p8802}, where Mars-analogs are scattered out and
avoid further embryo-embryo accretion events after a few million
years. The differences in these two evolutionary paths would lie
primarily in the accretion profiles of Earth and Venus-analogs that
alternately have long quiescent periods of accretion compared to
somewhat regular embryo accretion events found in more classical
models. In this way the systems in this work evolve similarly to that
found in \citet{Levison:2015p20168} - whereby that work forms $\sim$15
planetary embryos between 0.7-1.5~au. This system of embryos is built
rapidly and are stable for $\sim$10~Myr before embryo-embryo
collisions begin.

Finally, this work was primarily an exploration of the dynamics of
these systems as they grow from planetesimal to planets with and
without the affects of a gaseous nebula. These tests were ideal first
test cases for {\tt LIPAD} as the small annulus required less total
$N$ than will the upcoming full-disk terrestrial planet simulations
using the same code. However, the work found that for this test case
that the collisional aspects of the code are much less important than
is the total impact caused by the presence and lasting nature of the
gaseous solar nebula. Upcoming work will expand on this work using
similar techniques to study growth in a full-disk (0.7--3.0~au) and
scenarios including giant planet migration.

\section{Acknowledgements}

The participation of KJW and HL was supported by NASA’s SSERVI program
(Institute of the Science of Exploration Targets) through institute
grant number NNA14AB03A. This work used the Extreme Science and
Engineering Discovery Environment (XSEDE), which is supported by
National Science Foundation grant number ACI-1053575.

\clearpage

\bibliography{biblio}

\begin{thebibliography}{}
\expandafter\ifx\csname natexlab\endcsname\relax\def\natexlab#1{#1}\fi

\bibitem[{Carter {et~al.}(2015)Carter, Leinhardt, Elliott, Walter, \&
  Stewart}]{Carter:2015p19516}
Carter, P.~J., Leinhardt, Z.~M., Elliott, T., Walter, M.~J., \& Stewart, S.~T.
  2015, eprint arXiv, 1509, 7504

\bibitem[{Chambers(2001)}]{Chambers:2001p7618}
Chambers, J.~E. 2001, Icarus, 152, 205, (c) 2001: Academic Press

\bibitem[{Chambers(2013)}]{Chambers:2013p19990}
---. 2013, Icarus, 224, 43, (c) 2013 Elsevier Inc.

\bibitem[{Dauphas \& Pourmand(2011)}]{Dauphas:2011p19768}
Dauphas, N., \& Pourmand, A. 2011, Nature, 473, 489, (c) 2011: Nature

\bibitem[{Duncan {et~al.}(1998)Duncan, Levison, \& Lee}]{Duncan:1998p7713}
Duncan, M.~J., Levison, H.~F., \& Lee, M.~H. 1998, The Astronomical Journal,
  116, 2067, (c) 1998: The American Astronomical Society

\bibitem[{Dwyer {et~al.}(2015)Dwyer, Nimmo, \& Chambers}]{Dwyer:2015p20491}
Dwyer, C.~A., Nimmo, F., \& Chambers, J.~E. 2015, Icarus, 245, 145

\bibitem[{Fischer {et~al.}(2014)Fischer, Ciesla, \&
  Campbell}]{Fischer:2014p20911}
Fischer, R.~A., Ciesla, F., \& Campbell, A.~J. 2014, American Geophysical
  Union, 44

\bibitem[{Hansen(2009)}]{Hansen:2009p8802}
Hansen, B. M.~S. 2009, The Astrophysical Journal, 703, 1131

\bibitem[{{Ida} \& {Makino}(1993)}]{1993Icar..106..210I}
{Ida}, S., \& {Makino}, J. 1993, Icarus, 106, 210

\bibitem[{{Iwasaki} {et~al.}(2001){Iwasaki}, {Emori}, {Tanaka}, \&
  {Nakazawa}}]{2001DPS....33.1507I}
{Iwasaki}, K., {Emori}, H., {Tanaka}, H., \& {Nakazawa}, K. 2001, in Bulletin
  of the American Astronomical Society, Vol.~33, AAS/Division for Planetary
  Sciences Meeting Abstracts \#33, 1060

\bibitem[{Izidoro {et~al.}(2014)Izidoro, Haghighipour, Winter, \&
  Tsuchida}]{Izidoro:2014p15200}
Izidoro, A., Haghighipour, N., Winter, O.~C., \& Tsuchida, M. 2014, The
  Astrophysical Journal, 782, 31

\bibitem[{Jacobson \& Morbidelli(2014)}]{Jacobson:2014p18340}
Jacobson, S.~A., \& Morbidelli, A. 2014, Phil. Trans. R. Soc. A, 372, 0174

\bibitem[{Jin {et~al.}(2008)Jin, Arnett, Sui, \& Wang}]{Jin:2008p20003}
Jin, L., Arnett, W.~D., Sui, N., \& Wang, X. 2008, The Astrophysical Journal,
  674, L105

\bibitem[{Kenyon \& Bromley(2006)}]{Kenyon:2006p11683}
Kenyon, S., \& Bromley, B. 2006, The Astronomical Journal

\bibitem[{Kleine {et~al.}(2004)Kleine, Mezger, Palme, \&
  M{\"u}nker}]{Kleine:2004p19341}
Kleine, T., Mezger, K., Palme, H., \& M{\"u}nker, C. 2004, Earth and Planetary
  Science Letters, 228, 109

\bibitem[{Kleine {et~al.}(2009)Kleine, Touboul, Bourdon, Nimmo, Mezger, Palme,
  Jacobsen, Yin, \& Halliday}]{Kleine:2009p9784}
Kleine, T., Touboul, M., Bourdon, B., {et~al.} 2009, Geochimica et Cosmochimica
  Acta, 73, 5150

\bibitem[{Kokubo \& Genda(2010)}]{Kokubo:2010p9520}
Kokubo, E., \& Genda, H. 2010, arXiv, astro-ph.EP, 1003.4384v1, 21 pages, 3
  figures, 2 tables, ApJL in press

\bibitem[{Kokubo \& Ida(1998)}]{Kokubo:1998p9706}
Kokubo, E., \& Ida, S. 1998, Icarus, 131, 171

\bibitem[{{Kominami} \& {Ida}(2002)}]{2002Icar..157...43K}
{Kominami}, J., \& {Ida}, S. 2002, Icarus, 157, 43

\bibitem[{Leinhardt {et~al.}(2009)Leinhardt, Richardson, Lufkin, \&
  Haseltine}]{Leinhardt:2009p10318}
Leinhardt, Z.~M., Richardson, D.~C., Lufkin, G., \& Haseltine, J. 2009, Monthly
  Notices of the Royal Astronomical Society, 396, 718

\bibitem[{Levison {et~al.}(2012)Levison, Duncan, \&
  Thommes}]{Levison:2012p12338}
Levison, H., Duncan, M., \& Thommes, E. 2012, The Astronomical Journal

\bibitem[{Levison {et~al.}(2015)Levison, Kretke, Walsh, \&
  Bottke}]{Levison:2015p20168}
Levison, H.~F., Kretke, K., Walsh, K.~J., \& Bottke, W.~F. 2015, Proceedings of
  the National Academy of Sciences, 112, 46

\bibitem[{Morbidelli {et~al.}(2012)Morbidelli, Lunine, O'brien, Raymond, \&
  Walsh}]{Morbidelli:2012p11505}
Morbidelli, A., Lunine, J., O'brien, D., Raymond, S., \& Walsh, K. 2012, Annu.
  Rev. Earth. Planet. Sci., 40, 251, d

\bibitem[{Morishima(2015)}]{Morishima:2015p19487}
Morishima, R. 2015, arXiv, astro-ph.EP, 1508.07377v1

\bibitem[{Nimmo \& Kleine(2007)}]{Nimmo:2007p11241}
Nimmo, F., \& Kleine, T. 2007, Icarus, 191, 497

\bibitem[{O'brien {et~al.}(2006)O'brien, Morbidelli, \&
  Levison}]{Obrien:2006p8571}
O'brien, D.~P., Morbidelli, A., \& Levison, H.~F. 2006, Icarus, 184, 39

\bibitem[{O'brien {et~al.}(2014)O'brien, Walsh, Morbidelli, Raymond, \&
  Mandell}]{Obrien:2014p13867}
O'brien, D.~P., Walsh, K.~J., Morbidelli, A., Raymond, S.~N., \& Mandell, A.~M.
  2014, Icarus, 239, 74, (c) 2014 The Authors

\bibitem[{Ormel \& Dullemond{\ldots}(2010)}]{Ormel:2010p11708}
Ormel, C., \& Dullemond{\ldots}, C. 2010, The Astrophysical Journal {\ldots}

\bibitem[{Raymond {et~al.}(2009)Raymond, O'brien, Morbidelli, \&
  Kaib}]{Raymond:2009p11530}
Raymond, S.~N., O'brien, D.~P., Morbidelli, A., \& Kaib, N.~A. 2009, Icarus,
  203, 644

\bibitem[{{Thommes} {et~al.}(2003){Thommes}, {Duncan}, \&
  {Levison}}]{2003Icar..161..431T}
{Thommes}, E.~W., {Duncan}, M.~J., \& {Levison}, H.~F. 2003, Icarus, 161, 431

\bibitem[{Walsh {et~al.}(2011)Walsh, Morbidelli, Raymond, O'brien, \&
  Mandell}]{Walsh:2011p12463}
Walsh, K., Morbidelli, A., Raymond, S., O'brien, D., \& Mandell, A. 2011,
  Nature, 475, 206

\bibitem[{Wisdom \& Holman(1991)}]{Wisdom:1991p456}
Wisdom, J., \& Holman, M. 1991, Astronomical Journal (ISSN 0004-6256), 102,
  1528

\end{thebibliography}
\bibliographystyle{aasjournal}

\end{document}